\documentclass[preprint, showpacs,superscriptaddress,preprintnumbers,amsmath,amssymb]{revtex4}

\usepackage[dvipdfmx]{graphicx}
\usepackage{dcolumn}
\usepackage{bm}

\newcommand\+{\dagger}  
\newcommand\diag{\operatorname{diag}}

\newcommand\e{\mathrm{e}}

\newcommand\N{{\cal N}}

\newcommand{\bra}[1]{\langle {#1} |}
\newcommand{\ket}[1]{| {#1} \rangle}

\usepackage[normalem]{ulem}  
\usepackage[dvips]{color} 

\renewcommand\sout{\bgroup \color{red} \ULdepth=-.5ex \ULset}

\makeatletter
\renewcommand{\theequation}{\arabic{section}.\arabic{equation}}
\makeatother

\begin{document}

\begin{flushright}
TKYNT-12-02,\; KEK-TH-1523
\end{flushright}

\title{
Non-Abelian statistics of vortices\\ with multiple Majorana fermions 
}

\author{Yuji Hirono}
 \email{hirono@nt.phys.s.u-tokyo.ac.jp}
\affiliation{
Department of Physics, University of Tokyo, 
Hongo~7-3-1, Bunkyo-ku, Tokyo 113-0033, Japan
}

\author{Shigehiro Yasui}
\email{shigehiro.yasui@kek.jp}
\affiliation{KEK Theory Center, IPNS, KEK, 
1-1 Oho, Tsukuba, Ibaraki 305-0801, Japan}

\author{Kazunori Itakura} 
\email{kazunori.itakura@kek.jp}
\affiliation{KEK Theory Center, IPNS, KEK, 
1-1 Oho, Tsukuba, Ibaraki 305-0801, Japan}

\author{Muneto Nitta}
\email{nitta@phys-h.keio.ac.jp}
\affiliation{Department of Physics, and Research and Education Center 
for Natural Sciences, Keio University, 4-1-1 Hiyoshi, Yokohama, 
Kanagawa 223-8521, Japan}

\date{\today}

\begin{abstract}
We consider the exchange statistics of vortices, each of which 
traps an odd number ($N$) of Majorana fermions. We assume that 
the fermions in a vortex transform in
the vector representation of the $SO(N)$ group. 
Exchange of two vortices turns out to be non-Abelian, and the 
corresponding operator is further decomposed into two parts: 
a part that is essentially equivalent to the exchange operator 
of vortices having a single Majorana fermion in each vortex, 
and a part representing the Coxeter group. Similar decomposition 
was already found in the case with $N=3$, and the result shown 
here is a generalization to the case with an arbitrary odd $N$. 
We can obtain the matrix representation of the exchange 
operators in the Hilbert space that is constructed by using 
Dirac fermions non-locally defined by Majorana fermions trapped 
in separated vortices. We also show that the decomposition of 
the exchange operator implies tensor product structure in 
its matrix representation. 
\end{abstract}



\maketitle

\section{Introduction}

There has been considerable interest recently in 
zero-energy fermion modes trapped inside vortices 
in superconductors \cite{Jackiw:1981ee}.
Vortices in a chiral $p$-wave superconductor are endowed with 
non-Abelian statistics \cite{ReadGreen:00,Ivanov:2001} because of the
zero-energy Majorana fermions inside them \cite{Volovik:99}.
Excitations which obey non-Abelian statistics are called non-Abelian anyons.
They are expected to form the basis of topological quantum
computations \cite{Kitaev:2006,Nayak:2008zza} 
and have been investigated intensively \cite{NAstatistics}.
A recent classification of topological superconductors 
clarifies the condition that vortices have zero-energy 
Majorana (or Dirac) fermions in their cores \cite{SchnyderRFL:08,
Roy:2010}. 
One remarkable development is the fact that non-Abelian anyons 
in three dimensions \cite{Teo:2009qv} can be realized by monopoles 
with Majorana fermions trapped inside their cores,
and give a new non-Abelian statistics, the projective ribbon 
permutation statistics \cite{Freedman:2010ak}.

More recently, the non-Abelian statistics of vortices 
with additional $SO(3)$ symmetry has been investigated 
 and shown to have a novel structure \cite{Yasui:2010yh}.
In this case, the Majorana fermions form the vector representation 
of the $SO(3)$ group. 
It is shown that the representation under the exchange of two 
vortices is written by the tensor product of two matrices. 
One matrix is identical to the exchange matrix for vortices 
with a single Majorana fermion in each core, found by Ivanov 
\cite{Ivanov:2001} modulo trivial change of basis, 
and the other matrix is shown to be a generator of the Coxeter 
group, which is a symmetry group of certain polytopes in high 
dimensions \cite{Coxeter}. 

Such vortices can be physically realized in a ``color'' superconductor.
At extremely high densities and low temperatures, which could be 
achieved in the cores of compact stars, hadronic matter undergoes 
phase transition into quark matter, and it is expected to 
exhibit the color superconductivity \cite{Alford:2007xm}.
Unlike the ordinary superconductivity in a metal, color 
superconductivity is induced by condensation of diquarks,
which are pairs of two quarks (fermions having ``colors'' and
``flavors''). 
The order parameter of the color-flavor locked (CFL) phase is given by a  $3 \times 3$
matrix,
\begin{equation}
 \Phi_{\alpha i} = \epsilon_{\alpha \beta \gamma}
\epsilon_{ijk} \langle
\left(q^T\right)^j_\beta C\gamma_5 \left(q\right)^k_\gamma
  \rangle,
\end{equation}
where $q$ is the quark field, $\alpha, \beta, \gamma = {\rm r,g,b} \ 
(i,j,k = {\rm u,d,s})
$ 
are color (flavor) indices, and the transpose is employed with respect to
the spinor index.
At asymptotically high densities, the ground states are expected to be in
the CFL phase, in which the diquark acquires an expectation
value like $\Phi_{\alpha i} = \Delta \delta_{\alpha i}$, where $\Delta$
is a BCS gap function.
The original color $SU(3)_{\rm C}$ and flavor $SU(3)_{\rm F}$ symmetry
break down to an $SU(3)_{\rm C+F}$ symmetry, the elements of which are given by
``locked'' rotations of color and flavor, $\Phi \rightarrow U \Phi
U^{-1}$.
This structure is similar to the Balian-Werthamerer (BW) state of the ${}^3 {\rm He}$
superfluids, in which the order parameter is invariant under the locked
rotations of spin and orbit states.
We can argue that there exist topologically 
(and dynamically) stable vortices \cite{Balachandran:2005ev}
in the CFL phase, by examining the symmetry-breaking pattern.
The vortices in the CFL phase are color flux tubes as well as superfluid
vortices, since both
local and global symmetries are broken in the ground state.
In the presence of a vortex, the order parameter takes the
value like $\Phi(r) = \Delta \diag\{ f(r) e^{i \theta},g(r),g(r) \}$
where $r$ is the radial coordinate and $f(r)$ and $g(r)$ are functions
of $r$.
This kind of vortex solution breaks the $SU(3)_{\rm C+F}$ symmetry down to 
$SU(2)_{\rm C+F} \times U(1)$ symmetry inside the core
\cite{Nakano:2007dr,Eto:2009kg,Eto:2009bh,Eto:2009tr,Hirono:2010gq,Eto:2011mk}
and fermion zero modes belong to representations of $SU(2)_{\rm C+F}$.
It has been shown\cite{Yasui:2010yw,Fujiwara:2011za} that one CFL 
vortex has triplet and singlet Majorana fermions inside it \footnote{
The singlet Majorana fermion found in \cite{Yasui:2010yw} 
was in fact shown to diverge at the origin and consequently 
to be non-normalizable \cite{Fujiwara:2011za}.}.
Thus vortices in a color superconductor provide 
an example of the system with 
fermion zero modes in the vector representation of $SO(3)$ in their
cores, since the triplet of $SU(2)$ is equivalent to the vector
representation of $SO(3)$.
However, it should be emphasized that the results obtained in 
Ref.~\cite{Yasui:2010yh} do not depend on details of specific models. 
The only assumption adopted there is that a vortex
has Majorana fermions which transform in the vector representation of
$SO(3)$, and therefore we expect that such a system can be found 
in condensed matter systems such as exotic superconductors or 
ultra-cold atomic gasses. 

In this article, we generalize the results of Ref.~\cite{Yasui:2010yh}
obtained for $SO(3)$ to the case of $SO(N)$ where $N$ is an arbitrary 
odd integer $N\ge 3$. 
We discuss the exchange statistics of two vortices 
having $N$ Majorana fermions trapped in their cores, 
which are transformed in an $SO(N)$ group. 
Notice that the $SO(N)$ symmetry is the maximum symmetry 
in the presence of $N$ Majorana fermions which are real fields.
We discuss the exchange statistics in both the operator and 
representation levels. 
We find that both the operator and the matrix representation 
of the exchange operation 
generally have factorized structures for arbitrary odd $N$.
In particular, the matrix representation is written as a tensor product 
of two matrices, as previously found in the $SO(3)$ case. 
One of the two matrices is the one discussed in Ref.~\cite{Ivanov:2001} 
in the case that a single fermion is trapped in each vortex.
We show that the other is again a generator of the Coxeter group, 
as in the case of $SO(3)$. 
When one Majorana fermion is topologically protected 
in the vortex core such as in class-D topological superconductors, 
it is robust against perturbations and remains zero energy under the 
exchange operation. 
In addition to that, $N$ Majorana fermions remain zero energy 
as far as the $SO(N)$ symmetry remains intact.

This paper is organized as follows.
In Sec.~\ref{sec:single}, we review the non-Abelian statistics 
in the case of a single Majorana fermion studied in Ref.~\cite{Ivanov:2001}.
In Sec.~\ref{sec:SO(N)}, after summarizing the previous work on 
$SO(3)$ case \cite{Yasui:2010yh}, we present generalization to the 
case of multiple $(N)$ Majorana fermions with $SO(N)$ symmetry. 
We explicitly show factorization of the exchange operators into 
the known part similar to the case with a single Majorana fermion,
and the part corresponding to the Coxeter group. 
We also show an interesting decomposition of the Majorana fermion 
operators, which clarifies the action of the Coxeter group on 
Majorana fermions.
In Sec.~\ref{sec:tensor} we discuss the relation between the
decomposition of the exchange operator and its matrix representation.
Section~\ref{sec:summary} is devoted to summary and discussion. 
In Appendix \ref{sec:Proof} we prove the decomposition theorem 
of the exchange operation of vortices.

\setcounter{equation}{0}
\section{Vortices with $N=1$ \label{sec:single}}

We briefly review how non-Abelian statistics emerges 
for a set of $n$ vortices, each of which contains a 
single Majorana fermion
in its core \cite{Ivanov:2001}. This provides 
the simplest example of non-Abelian statistics of vortices, 
but as we will see later we can always identify the same structure even
for the case with multiple Majorana fermions as far as the number of fermions 
$N$ is odd.

Consider an exchange of two vortices in a system of $n=2m$ vortices
\footnote{
The structure of the Hilbert space in the presence of an odd number
vortices is recently discussed in Ref.~\cite{Jackiw:2011tk}.
}. 
Each vortex has a single Majorana fermion localized at the core of it, and 
one can specify the position of a vortex on a two-dimensional plane (we 
label the vortices $k=1,\cdots,n$). 
Notice that the trapped Majorana fermion has zero energy, which gives rise 
to degeneracy of the ground (lowest energy) state. Since the existence of 
Majorana fermions are topologically guaranteed, the degeneracy is not 
disturbed by small perturbations, and hence we treat the exchange of 
vortices as an adiabatic process.

Let an operation $T_{k}$ be defined as an exchange of the $k$-th and 
$(k+1)$-th vortices, in which the $(k+1)$-th vortex turns around the 
$k$-th vortex in an anti-clockwise way. All the exchanges of two vortices 
are realized by successive application of the exchanges of adjacent 
vortices $T_k$, and they form a braid group $B_{n}$. 
They indeed satisfy the braid relations: 
\begin{eqnarray}
T_k T_\ell T_k &=& T_\ell T_k T_\ell  \quad {\rm for} \quad |k-\ell| = 1,
\label{eq:braid1} \\
T_k T_\ell  &=& T_\ell T_k  \quad {\rm for}  \quad |k-\ell| > 1.
\label{eq:braid2}
\end{eqnarray}

Recall that the vortices are accompanied by Majorana fermions, 
and one can express the action of $T_k$ on Majorana fermions
as a transformation. To this end, we define
a Majorana operator $\gamma_{k}$ corresponding to 
the Majorana fermion in the $k$-th vortex \cite{Ivanov:2001}, 
satisfying the self-conjugate condition $(\gamma_{k})^{\dag} = \gamma_{k}$ 
and the anti-commutation relation 
$\{ \gamma_{k}, \gamma_{\ell} \}=2\delta_{k\ell}$ (the Clifford algebra).
Then, the operation $T_{k}$ induces the following transformation 
\cite{Ivanov:2001}:
\begin{eqnarray}
T_{k} :
\left\{
\begin{array}{l}
 \gamma_{k} \rightarrow \gamma_{k+1} \\
 \gamma_{k+1} \rightarrow -\gamma_{k}
\end{array}
\right. ,
\label{eq:Tk_single}
\end{eqnarray}
with the rest $\gamma_{\ell}$ ($\ell \neq k$, $k+1$) unchanged.
One can explicitly check that the transformation (\ref{eq:Tk_single}) 
satisfies the braid relations (\ref{eq:braid1}) and (\ref{eq:braid2}).
The minus sign in the second line is essential for the non-Abelian statistics
because it gives $T_k^4=1$ (the Bose-Einstein or 
Fermi-Dirac statistics give just $T_k^2=1$).
The transformation (\ref{eq:Tk_single}) is realized by the unitary operator
\begin{eqnarray}
\tau_{k} \equiv \exp \left(\frac{\pi}{4}\gamma_{k+1}\gamma_k \right)=
\frac{1}{\sqrt{2}} \left(1+\gamma_{k+1}\gamma_{k}\right)\, .
\label{eq:tau-k}
\end{eqnarray}
Indeed, one finds 
\begin{eqnarray}
&& \tau_{k} \gamma_{k} \tau_{k}^{-1}  = \gamma_{k+1}\, ,
\label{eq:Ivanov_transf_1} \\
&& \tau_{k} \gamma_{k+1} \tau_{k}^{-1} = -\gamma_{k}\, , 
\label{eq:Ivanov_transf_2}\\
&& \tau_{k} \gamma_{\ell} \tau_{k}^{-1} = \gamma_{\ell} 
\;\; \ (\ell \neq k,k+1)\, .
\label{eq:Ivanov_transf_3}
\end{eqnarray}
We call the operator $\tau_{k}$ for this single 
Majorana case the `Ivanov operator' since it was first found by 
Ivanov \cite{Ivanov:2001}. One can explicitly see that 
this transformation is indeed non-Abelian in the matrix representation 
of $\tau_k$. To construct the Hilbert space on which the operator 
$\tau_k$ acts, we define 
Dirac fermions \footnote{We use $k,\ell=1,\cdots,n$ to label the vortices and 
the trapped Majorana fermions, while $K,L=1,\cdots, n/2=m$ to label the Dirac
fermions which are constructed from two Majorana fermions.} 
by using two Majorana fermions at adjacent vortices
$\Psi_{K} = ( \gamma_{2K-1} + i \gamma_{2K} )/2$ with $K=1,\cdots,m$.
These Dirac fermion operators satisfy the usual anti-commutation 
relations, 
\begin{eqnarray}
\{ \Psi^{}_{K}, \Psi_{L}^{\dag} \}=\delta_{KL}, \quad 
\{ \Psi_{K}, \Psi_{L} \} = \{ \Psi_{K}^{\dag}, \Psi_{L}^{\dag} \} = 0.
\end{eqnarray}
If one defines $\Psi_K$ and $\Psi^\dag_K$ as the annihilation and 
creation operators, respectively, then one 
can construct the Hilbert space by acting the creation operators 
$\Psi_{K}^{\dag}$'s on the ``Fock vacuum-state'' $\ket{0}$ 
defined by $\Psi_{K} \ket{0}=0$ for all $K$. 
Within this Hilbert space, the operators $\tau_k$'s are 
now expressed as matrices which we call the Ivanov matrices.
The Ivanov matrices contain off-diagonal elements representing 
the non-Abelian statistics. 
Explicit forms of the Ivanov matrices 
for two and four vortices are given in Ref.~\cite{Ivanov:2001}, and 
also in Ref.~\cite{Yasui:2010yh} with a 
different expression, both of which are 
related to each other by unitary transformations.

\setcounter{equation}{0}
\section{Vortices with multiple Majorana fermions $N\ge 3$ \label{sec:SO(N)}}

Now let us turn to the case with vortices having multiple 
Majorana fermions in their cores. Each vortex traps
$N$ Majorana fermions $\gamma_k^a$ $(a=1,\cdots,N)$ which
transform in the vector
representation of $SO(N)$ symmetry, hence we call them non-Abelian
Majorana fermions. We consider $N$ to be an arbitrary odd number, and
this is a generalization of the simplest non-trivial case $N=3$ in
Ref.~\cite{Yasui:2010yh}. 

The non-Abelian Majorana operators $\gamma_k^a$ satisfy the self-conjugate 
conditions and the anti-commutation relations: 
\begin{align}
(\gamma^a_k)^\+ =\gamma^a_k , \quad
\{\gamma^a_k, \gamma^b_\ell \} = 2 \delta^{ab}\delta_{k\ell}\, .
\end{align}
We define the exchange of the $k$-th and $(k+1)$-th vortices 
so that the Majorana fermion operator with each $a$ transforms 
in the same way as the case with a single Majorana 
fermion (see Eq.~(\ref{eq:Tk_single})) \footnote{One may allow for 
mixture of indices $a$ under the exchange of two vortices, but here we discuss 
the simplest case where such mixing does not occur.}: 
\begin{eqnarray}
T_k : \left\{ 
\begin{array}{l}
 \gamma_{k}^{a}\quad \rightarrow\, \gamma_{k+1}^{a} \\
 \gamma_{k+1}^{a} \rightarrow -\gamma_{k}^{a}  
\end{array}
\right.  \qquad {\rm for \ all}\ a\, ,
\label{eq:exchange_Majorana}
\end{eqnarray}
with the rest $\gamma_\ell^a$ ($\ell \neq k$, $k+1$) unchanged. 
The exchange operations $T_{k}$'s satisfy the braid 
relations (\ref{eq:braid1}) and (\ref{eq:braid2}).
Their action on $\gamma^a_\ell$'s can be represented in terms of
non-Abelian Majorana operators $\gamma_{k}^{a}$'s in the following way.
Since the transformation (\ref{eq:exchange_Majorana}) for each $a$
is equivalent to the single Majorana case, one can use the same expression
for the unitary operator which induces the transformation for each $a$:
\begin{equation}
 \tau^a_k \equiv \exp\left( \frac{\pi}{4} \gamma^a_{k+1}\gamma^a_k \right) 
= \frac{1}{\sqrt{2}}\left( 1 + \gamma^a_{k+1} \gamma^a_k \right)\, .
\end{equation}
Thus, the exchange operator for the vortices having multiple fermions 
should be represented as the product of them:
\begin{equation}
 \tau_{k}^{[N]} \equiv \prod_{a=1}^{N} \tau^a_k \, . \label{eq:exchange-op}
\end{equation}
This exchange operator is $SO(N)$ invariant as shown in the next section. 
One can check that the operator $\tau_{k}^{[N]}$ applied to 
$\gamma^a_\ell$
indeed generates the desired transformation (\ref{eq:exchange_Majorana}):
\begin{eqnarray}
&& \tau_{k}^{[N]} \gamma^a_{k} (\tau_{k}^{[N]})^{-1} = \gamma^a_{k+1}\, , \\
&& \tau_{k}^{[N]} \gamma^a_{k+1} (\tau_{k}^{[N]})^{-1} = -\gamma^a_{k}\, , \\
&& \tau_{k}^{[N]} \gamma^a_{\ell} (\tau_{k}^{[N]})^{-1}=\gamma^a_{\ell} \qquad 
(\ell \neq k,\, k+1)\, .
\end{eqnarray}
This transformation is again non-Abelian, which is 
explicitly seen in the matrix representation. 

To obtain the matrix representation, we can perform the same 
procedure as in the case with single Majorana fermions. 
Namely, by defining the Dirac fermion operators 
\begin{equation}
 \Psi^a_K \equiv \frac12 ( \gamma^a_{2K-1} + i \gamma^a_{2K} ) ,\qquad 
 \Psi^{a\dagger}_K \equiv \frac12( \gamma^a_{2K-1} - i \gamma^a_{2K} ) 
 \label{eq:Dirac}
\end{equation}
which satisfy $(K,L=1,\cdots,m)$ 
\begin{eqnarray}
\{ \Psi_{K}^a, \Psi_{L}^{b\dag} \}=\delta_{KL}\delta^{ab}, \quad 
\{ \Psi_{K}^a, \Psi_{L}^b \} = \{ \Psi_{K}^{a\dag}, \Psi_{L}^{b\dag} \} = 0, 
\end{eqnarray}
we can construct the Hilbert space. 
Then, we can find matrix representation of $\tau_k^{[N]}$.
In Ref.~\cite{Yasui:2010yh}, three of us explicitly constructed 
the Hilbert space for the case of $N=3$ and $n=2,4$, 
and found matrix expression of $\tau_k^{[3]}$
according to different representations of $SO(3)$. The matrices have 
off-diagonal elements and thus they are non-Abelian. In principle, one 
can do the same thing for an arbitrary odd $N$.
In the present paper, however, we first
look into interesting structure of the operator $\tau_k^{[N]}$, 
which was also suggested in the previous paper \cite{Yasui:2010yh}. 
Namely, the operator $\tau_k^{[N]}$ itself can be decomposed into 
two parts. Then, we discuss the relation between the decomposition of 
$\tau_k^{[N]}$ and the matrix representation of 
$\tau_k^{[N]}$ in Sec.~\ref{sec:tensor}.

\setcounter{equation}{0}
\section{The Coxeter group for multiple Majorana fermions $N\ge3$}

In the previous analysis \cite{Yasui:2010yh} with $N=3$, 
we found that the 
matrix representation of $\tau_k^{[3]}$ in the four-vortex sector can 
be decomposed into a tensor product of two matrices: one is the same 
as the Ivanov matrix for the single Majorana fermion case, and the 
other is identified with generators of the Coxeter group. We also 
found that the similar decomposition 
is possible at the operator level \cite{Yasui:2010yh}. 
Namely, one can express $\tau_k^{[3]}$
as a product of two distinct operators 
which give rise to the corresponding
matrices in the matrix representation. 
In this section, we discuss that 
such a decomposition at the operator level holds even for an 
arbitrary odd number of $N$.

Before we go into details, let us define some useful notations.
We first define a composite operator $\Gamma^a_k$ by 
\begin{equation}
 \Gamma^a_k \equiv \gamma^a_{k+1} \gamma^a_{k}\, ,
\end{equation}
which have the following properties,
\begin{eqnarray}
\{ \Gamma^a_k, \Gamma^b_l \} & = & -2 \delta_{kl} \quad 
{\rm for} \quad  a = b{\;\;}{\rm and}{\;\;} |k-l| \leq 1 \,, \\
\left[ \Gamma^a_k, \Gamma^b_l \right] & = & 0 
\quad \quad \quad {\rm for} \quad a \neq b,{\;\;} {\rm or}{\;\;} (a=b {\;\;} {\rm
and} {\;\;} |k-l| > 1 ) \, .
\label{eq:prop-2}
\end{eqnarray}
For later convenience, we define another composite operator $(1 \le n \le N)$
\begin{equation}
\Gamma^{(n)}_k \equiv
\frac{1}{(n!)^2}\, 
\e^{\frac{\pi}{2}i(n-1)}
\epsilon^{a_1\cdots a_N}\epsilon^{b_1\cdots b_N} \, 
\delta^{a_{n+1}}_{b_{n+1}}\cdots \delta^{a_{N}}_{b_{N}}\, 
\gamma_{k+1}^{a_1}\cdots \gamma_{k+1}^{a_n} \, 
\gamma_{k}^{b_1}\cdots \gamma_{k}^{b_n} \, , 
\label{Gamma_n}
\end{equation}
where $\epsilon^{a_1\cdots a_N}$ is the completely antisymmetric tensor. 
It is evident that the operators $\Gamma_k^{(n)}$'s
are $SO(N)$ invariants for all $n$.
For example, for $N = 3$, 
\begin{equation}
 \Gamma^{(1)}_k = \Gamma^1_k +  \Gamma^2_k + \Gamma^3_k, 
\quad
 \Gamma^{(2)}_k = 
\Gamma^{1}_k\Gamma^2_k
+\Gamma^{2}_k\Gamma^3_k
+\Gamma^{3}_k\Gamma^1_k 
, \quad 
 \Gamma^{(3)}_k = \Gamma^1_k\Gamma^2_k\Gamma^{3}_k\, . \label{N=3}
\end{equation}
%
With those composite operators, the exchange operator $\tau_{k}^{[N]}$ 
can be represented as 
\begin{equation}
 \tau_{k}^{[N]} = 
\left( \frac{1}{\sqrt{2}} \right)^{\!\! N} \, \prod_{a=1}^{N} 
\left(1+\Gamma^a_k \right)
=\exp \left\{\frac{\pi}{4}\sum_a\Gamma_k^a\right\}
= \left( \frac{1}{\sqrt{2}}\right)^{N}
 \sum_{n=1}^{N}  \Gamma^{(n)}_k.
\label{tau_new}
\end{equation}
In the final expression, we confirm that the operator $\tau_{k}^{[N]}$ is 
$SO(N)$-invariant.

\subsection{The case of $SO(3)$}

The simplest nontrivial case with $N=3$ provides us with useful 
information which is helpful in discussing the decomposition for the general 
case $N$. Let us first recall that the operator $\tau_{k}^{[N=3]}$ has been 
found to be decomposed into two parts (see Appendix B in 
Ref.~\cite{Yasui:2010yh}):
\begin{equation}
\tau_{k}^{[3]} = \sigma_{k}^{[3]} h_{k}^{[3]}\, ,
\end{equation}
where both of the operators $\sigma_{k}^{[3]}$ and $h_{k}^{[3]}$ are 
given in terms of the Majorana operators $\gamma_k^a$ as 
\begin{eqnarray}
&&{\sigma}_{k}^{[3]} 
  = \frac{1}{2} 
    \Big( 1 - \gamma_{k+1}^{1}\gamma_{k+1}^{2}\gamma_{k}^{1}\gamma_{k}^{2} 
            - \gamma_{k+1}^{2}\gamma_{k+1}^{3}\gamma_{k}^{2}\gamma_{k}^{3} 
- \gamma_{k+1}^{3}\gamma_{k+1}^{1}\gamma_{k}^{3}\gamma_{k}^{1}
     \Big) \, ,
\end{eqnarray}
and
\begin{eqnarray}
{h}_{k}^{[3]} 
  = \frac{1}{\sqrt{2}} 
    \Big( 1- \gamma_{k+1}^{1}\gamma_{k+1}^{2}\gamma_{k+1}^{3}
             \gamma_{k}^{1} \gamma_{k}^{2} \gamma_{k}^{3}
    \Big)\, .
\end{eqnarray}
One can also rewrite them compactly in new notations as 
(see Eqs.~(\ref{Gamma_n}) and (\ref{N=3}))
\begin{equation}
 \sigma^{[3]}_k = 
 \frac{1}{2}\left( 1 + \Gamma^{(2)}_k \right),
 \quad 
h^{[3]}_k =
\frac{1}{\sqrt{2}}
\left( 1+ \Gamma^{(3)}_k \right) .
\end{equation}
Note that $\sigma_{k}^{[3]}$ and $h_{\ell}^{[3]}$ are commutative 
for any pair of $k$ and $\ell$.
Thus, $\tau_{k}^{[3]}$ is the product of $\sigma_{k}^{[3]}$ and 
$h_{k}^{[3]}$. It was shown that the operators $ \sigma^{[3]}_k $'s 
satisfy the Coxeter relations
\footnote{A Coxeter group $S$ is defined as a group with distinct
generators $s_i \in S$ ($i=1,2,3,\cdots$) satisfying the following 
two conditions \cite{Coxeter}: 
(a) $s_i^{2}=1$ and (b) $(s_i\,s_j)^{m_{i,j}}=1$ 
with a positive integer $m_{i,j} \ge 2$ for $i \neq j$.
It should be noted that condition (a) gives $m_{i,i}=1$.
Elements $m_{i,j}$ can be summarized as 
the Coxeter matrix $({\sf M})_{ij}=m_{i,j}$. 
In our case of the $n=2m$ vortices, the Coxeter matrix 
is given by a $(2m-1)\times(2m-1)$ matrix whose elements 
are 1 (diagonal elements, $m_{i,i}=1$), 
3 (adjacent elements, $m_{i,i+1}=m_{i+1,i}=3$) and 2 (all the others). 
},
\begin{eqnarray}
({\sigma}_{k}^{[3]})^{2} &=& 1, \label{eq:Coxeter3_1} \\
({\sigma}_{k}^{[3]} {\sigma}_{\ell}^{[3]})^{3} &=& 1 
\quad {\rm for}\quad |k-\ell|=1,
  \label{eq:Coxeter3_2} \\
({\sigma}_{k}^{[3]} {\sigma}_{\ell}^{[3]})^{2} &=& 1 
\quad {\rm for}\quad |k-\ell|>1.
\label{eq:Coxeter3_3}
\end{eqnarray}
The Coxeter group found here for the $n=2m$ vortices is classified 
as $A_{2m-1}$. It is also known as the symmetric group $S_{2m}$,
which is the symmetry group of a regular $(2m-1)$-simplex.

In contrast, the other part $h_{k}^{[3]}$ works in the same
way as the Ivanov operator $\tau_k$ defined in Eq.~(\ref{eq:tau-k}), 
although $h_k^{[3]}$ has a more complicated structure.
This is naturally understood if one 
notices that the operator $\gamma_{k}^{1} \gamma_{k}^{2} \gamma_{k}^{3}$ is 
$SO(3)$ invariant (singlet), and thus it treats three Majorana fermions 
together as if it does not have any $SO(3)$ index. This picture is very 
useful when we consider the general case with $N$.

\subsection{The case of $SO(N)$ with arbitrary odd $N$}

The previous analysis suggests that it is possible to decompose 
the full exchange operator $\tau_k^{[N]}$ into two parts. 
This is indeed the case. 
We find that the operator $\tau^{[N]}_k$ can be decomposed as
\begin{equation}
\begin{split}
 \tau^{[N]}_k 
&=
\sigma^{[N]}_k 
h^{[N]}_k\, ,
\end{split}
\label{eq:formula-n}
\end{equation}
where $\sigma^{[N]}_k $ and  $h^{[N]}_k $ are defined by 
using the notation introduced before as 
\begin{eqnarray}
\sigma^{[N]}_k 
&\equiv &
 \left( \frac{1}{\sqrt{2}}
\right)^{N-1}
\left(
1 + \Gamma^{(2)}_k + \Gamma^{(4)}_k + \dots + \Gamma^{(N-1)}_k
\right), \label{eq:def-sigma}\\
 h^{[N]}_k 
&\equiv&
 \frac{1}{\sqrt{2}} 
\left(
1+ \Gamma^{(N)}_k
\right). \label{eq:def-h}
\end{eqnarray}
Note that $\sigma_{k}^{[N]}$ and $h_{k}^{[N]}$ are $SO(N)$ invariant,
and $\sigma_{k}^{[N]}$ and $h_{\ell}^{[N]}$ are commutative for any $k$ and 
$\ell$.
One can readily verify the decomposition~(\ref{eq:formula-n}). 
By using 
Eqs.~(\ref{eq:def-sigma}) and (\ref{eq:def-h}), 
one can check that the product $\sigma^{[N]}_k h^{[N]}_k$ 
is equal to the last equation in Eq.~(\ref{tau_new}) if one uses
Eqs.~(\ref{eq:prop-1}) and (\ref{eq:prop-2}).

First of all, let us discuss the properties of $h_{k}^{[N]}$.
The analysis for the case $N=3$ suggests that if one treats 
multiple Majorana fermions in a vortex in a unit, then the action of 
$\tau_k^{[N]}$ will be essentially equivalent to the Ivanov operator. 
This motivates us to introduce the following 
``singlet Majorana operator'' locally defined on the $k$-th vortex:
\begin{equation}
 \overline{\gamma}_k \equiv 
\frac{1}{N!}
{\rm e}^{i\frac{\pi}{4}(N-1)} 
\epsilon^{a_1 a_2 \cdots a_N}
 \gamma_k^{a_1} \gamma_k^{a_2} \cdots \gamma_k^{a_N}\, ,
  \label{eq:composite}
\end{equation}
which is manifestly invariant under the $SO(N)$ transformation.
The phase factor is included so that the operator becomes 
self-conjugate $(\overline \gamma_k)^\dagger =\overline \gamma_k$, 
and satisfies 
the Clifford algebra $\{\overline{\gamma}_k, \overline{\gamma}_\ell \} 
= 2 \delta_{k\ell}$. 
Notice that these properties of $\overline\gamma_k$ are the same as those of a 
single Majorana operator. For $N=3$, one finds 
$\overline\gamma_k=i\gamma_{k}^{1} \gamma_{k}^{2} \gamma_{k}^{3}$, and the 
operator $h_k^{[3]}$ can be compactly expressed as 
$h_k^{[3]}=\frac{1}{\sqrt2}(1+\overline\gamma_{k+1} \overline\gamma_k)$, which has the 
same structure as the Ivanov operator (\ref{eq:tau-k}). 
For arbitrary odd $N$, we find that $h^{[N]}_k$ can also be expressed as
\begin{equation}
h^{[N]}_k 
= \exp\left(\frac{\pi}{4} \overline{\gamma}_{k+1}\overline{\gamma}_k\right)
= \frac{1} {\sqrt{2}} \left( 1 
               + \overline{\gamma}_{k+1}\overline{\gamma}_{k} \right)\,,
\label{eq:h_k^N}
\end{equation}
by noting the relation $\Gamma_{k}^{(N)}=\overline{\gamma}_{k+1} \overline{\gamma}_{k}$.
Then, $h^{[N]}_k$ works on $\overline{\gamma}_{\ell}$ as
\begin{eqnarray}
 && h_{k}^{[N]} \overline{\gamma}_{k} (h_{k}^{[N]})^{-1} = \overline{\gamma}_{k+1}, \label{eq:h-1} \\
 && h_{k}^{[N]} \overline{\gamma}_{k+1} (h_{k}^{[N]})^{-1} = -\overline{\gamma}_{k}, \label{eq:h-2} \\
 && h_{k}^{[N]} \overline{\gamma}_{\ell} (h_{k}^{[N]})^{-1}  = \overline{\gamma}_{\ell} \qquad 
(\ell \neq k,\, k+1)\, ,
\label{eq:h-3}
\end{eqnarray}
for an arbitrary odd $N\ge3$.
Interestingly, this is the same as the transformation
(\ref{eq:Ivanov_transf_1})~--~(\ref{eq:Ivanov_transf_3}) 
induced by the Ivanov operator with $N=1$.
Therefore, the operator $h^{[N]}_k$ for the singlet Majorana operator
$\overline{\gamma}_{k}$ is equivalent to the Ivanov operator $\tau_{k}$
for the single Majorana operator $\gamma_{k}$.

Next, we discuss the properties of the operators $\sigma_k^{[N]}$. 
Similarly to the $N=3$ case, $\sigma_{k}^{[N]}$ are generators 
of the Coxeter group.
Indeed, it can be confirmed that $\sigma_{k}^{[N]}$ satisfy (see 
Appendix \ref{sec:Proof} for details)
\begin{eqnarray}
({\sigma}_{k}^{[N]})^{2} &=& 1, \label{eq:Coxeter-A1} \\
({\sigma}_{k}^{[N]} {\sigma}_{\ell}^{[N]})^{3} &=& 1 \quad 
{\rm for}\quad |k-\ell|=1,
  \label{eq:Coxeter-A2} \\
({\sigma}_{k}^{[N]} {\sigma}_{\ell}^{[N]})^{2} &=& 1 \quad 
{\rm for}\quad |k-\ell|>1,
\label{eq:Coxeter-A3}
\end{eqnarray}
and these relations induce the same Coxeter group
for $n=2m$ vortices.
We thus have found that, for an arbitrary odd $N$, the operators 
$\sigma_{k}^{[N]}$'s again obey the Coxeter relations of $A_{2m-1}$.

In fact, one can ``derive'' the decomposition (\ref{eq:formula-n})
by first defining $h^{[N]}_k$ in terms of the singlet Majorana operators 
$\overline \gamma_k$ as in Eq.~(\ref{eq:h_k^N}), and then assuming the 
factorized form (\ref{eq:formula-n}). The operator 
$\sigma_{k}^{[N]} = \tau_{k}^{[N]} (h_{k}^{[N]})^{-1}$ thus obtained  
indeed coincides with Eq.~(\ref{eq:def-sigma}), and the commutativity 
of $\sigma_k^{[N]}$ and $h_\ell^{[N]}$ for any $k$ and $\ell$ confirms the
assumption of factorization.

We thus have found that, for an arbitrary odd $N$, the exchange operator  
$\tau_{k}^{[N]}$ is expressed as 
a product of a generator of the Coxeter group $A_{2m-1}$ (for the vortex 
number $n=2m$) and the Ivanov operator for a single Majorana fermion.

\subsection{Decomposition of the Majorana operators}
 
Let us recall that the exchange of the Majorana operators 
$\gamma_{\ell}^{a}$'s is originally defined as the operation $T_{k}$ in 
Eq.~(\ref{eq:exchange_Majorana}). It is not apparently clear how the 
decomposed structure of the operator $\tau_k^{[N]}$ indeed 
works in the exchange operation $T_k$. To understand this, it is instructive 
to notice that the Majorana operator $\gamma_{k}^{a}$ can be rewritten as
\begin{eqnarray}
\gamma_{k}^{a} = \widetilde{\gamma}_{k}^{a} \, \overline{\gamma}_{k}\, ,
\label{decomp_Majorana}
\end{eqnarray}
where $\widetilde{\gamma}_{k}^{a}$ is a composite operator in the vector
representation of $SO(N)$ defined locally on the $k$-th vortex as
\begin{eqnarray}
 \widetilde{\gamma}_{k}^{a} \equiv 
\frac{1}{(N-1)!}\, 
{\rm e}^{i\frac{\pi}{4}(N-1)}
  \epsilon^{a a_{1}\cdots a_{N-1}} \gamma_{k}^{a_1} \cdots 
  \gamma_{k}^{a_{N-1}},
\label{eq:composite2}
\end{eqnarray}
and $\overline{\gamma}_{k}$ is the singlet Majorana operator defined in
Eq.~(\ref{eq:composite}). The two operators 
$\widetilde{\gamma}_{k}^{a}$ and $\overline{\gamma}_{\ell}$ 
commute with each other for any pair of $k$ and $\ell$.
This expression allows us to extract, from the 
Majorana operator $\gamma_k^a$, the part of a singlet Majorana fermion 
$\overline \gamma_k$ whose properties are well understood. Notice that 
$\overline \gamma_k$ ($\widetilde \gamma_k^a$) is composed of an odd 
number $N$ (an even number $N-1$) of Majorana fermion operators.

Since $\widetilde \gamma_k^a$ and $\overline\gamma_k$ are 
expressed in terms of the original Majorana operator $\gamma_k^a$, 
one can immediately find how they are transformed in the exchange 
$T_k$.
Namely, we find the transformation of $\widetilde{\gamma}_{k}^{a}$ 
and $\overline{\gamma}_{k}$ by $T_{k}$ as
\begin{eqnarray}
T_k : \left\{ 
\begin{array}{l}
 \widetilde{\gamma}_{k}^{a}\quad \rightarrow\, \widetilde{\gamma}_{k+1}^{a} \\
 \widetilde{\gamma}_{k+1}^{a} \rightarrow \widetilde{\gamma}_{k}^{a}  
\end{array}
\right. , \quad {\rm for \ all}\ a\, ,
\label{eq:exchange_tilde}
\end{eqnarray}
without a minus sign, and 
\begin{eqnarray}
T_k : \left\{ 
\begin{array}{l}
 \overline{\gamma}_{k} \quad \rightarrow\, \overline{\gamma}_{k+1} \\
 \overline{\gamma}_{k+1} \rightarrow -\overline{\gamma}_{k}  
\end{array}
\right. , \quad {\rm for \ all}\ a\, ,
\label{eq:exchange_bar}
\end{eqnarray}
with a minus sign, while the rest $\widetilde{\gamma}_{\ell}^{a}$ 
and $\overline{\gamma}_{\ell}$ ($\ell \neq k$ and $k+1$) are unchanged
\footnote{In general, a composite operator made by an even (odd)
number of the Majorana operators is transformed as in 
Eq.~(\ref{eq:exchange_tilde}) (Eq.~(\ref{eq:exchange_bar})).}.
It is easily checked that the simultaneous transformation of 
$\widetilde{\gamma}_{k}^{a}$ and $\overline{\gamma}_{k}$ reproduces 
the transformation of $\gamma_{k}^{a}$ in Eq.~(\ref{eq:exchange_Majorana}).
Therefore, we observe from Eq.~(\ref{eq:exchange_tilde}) that
$\widetilde{\gamma}_{\ell}^{a}$'s are transformed by a symmetric group
$S_{2m}$, or the Coxeter group of $A_{2m-1}$ (for the vortex number
$n=2m$), and from Eq.~(\ref{eq:exchange_bar}) that $\overline{\gamma}_{\ell}$'s
are transformed as in the same way as the Ivanov operators for single Majorana 
fermions \footnote{\label{even-N}
If one considers the case when $N$ is an even number, one may define
a composite operator by $\overline{\gamma}_{k}' \equiv \frac{1}{N!}{\rm
e}^{i\frac{\pi}{4}N}  \gamma_k^1  \gamma_k^2 \cdots
\gamma_k^{N}$. The operator is self-conjugate 
$(\overline{\gamma}^\prime_{k})^{\dag} = \overline{\gamma}^\prime_{k}$, 
but  does not satisfy the Clifford algebra. 
Furthermore, since $N$ is even, the operation $T_{k}$ gives the
transformation, $\overline{\gamma}^\prime_{k}
\rightarrow \overline{\gamma}^\prime_{k+1}$, 
$\overline{\gamma}^\prime_{k+1} \rightarrow
\overline{\gamma}^\prime_{k}$ and the rest $\overline{\gamma}^\prime_{\ell}$ 
($\ell \neq k$ and $k+1$) unchanged. 
Hence the composite operator $\overline{\gamma}^\prime_{k}$ for even $N$ does 
not transform like a singlet Majorana fermion operator.}.

The exchange properties of $\widetilde{\gamma}_{k}^{a}$ and
$\overline{\gamma}_{k}$ in Eqs.~(\ref{eq:exchange_tilde}) and
(\ref{eq:exchange_bar}) can be discussed at the operator level. 
Because $\sigma_{k}^{[N]}$ and $\overline{\gamma}_{l}$ ($h_{k}^{[N]}$ and $\widetilde{\gamma}_{l}^{a}$) are commutative for any pair of $k$ and $\ell$,
\begin{eqnarray}
[ \sigma_{k}^{[N]},\overline{\gamma}_{\ell}]=
[ h_{k}^{[N]}, \widetilde{\gamma}_{\ell}^{a}]=0\, , \label{eq:commute}
\end{eqnarray}
the transformation $\tau_{k}^{[N]} \gamma_{\ell}^{a} (\tau_{k}^{[N]})^{-1}$
is decomposed as
\begin{eqnarray}
\tau_{k}^{[N]} \gamma_{\ell}^{a} (\tau_{k}^{[N]})^{-1} 
= \left\{ \sigma_{k}^{[N]} \widetilde{\gamma}_{\ell}^{a} 
(\sigma_{k}^{[N]})^{-1} \right\} 
\, \, \left\{ h_{k}^{[N]} \overline{\gamma}_{\ell} (h_{k}^{[N]})^{-1}\right\}\, .
\end{eqnarray}
Therefore, $\widetilde{\gamma}_{l}^{a}$ and $\overline{\gamma}_{l}$ 
are transformed by $\sigma_{k}^{[N]}$ and $h_{k}^{[N]}$, respectively.
From Eqs.~(\ref{eq:def-sigma}) and (\ref{eq:composite2}),
$\widetilde{\gamma}_{k}^{a}$ is transformed as
\begin{eqnarray}
 && \sigma_{k}^{[N]} \widetilde{\gamma}_{k}^{a} (\sigma_{k}^{[N]})^{-1} = \widetilde{\gamma}_{k+1}^{a}, \\
 && \sigma_{k}^{[N]} \widetilde{\gamma}_{k+1}^{a} (\sigma_{k}^{[N]})^{-1} = \widetilde{\gamma}_{k}^{a}, \\
 && \sigma_{k}^{[N]} \widetilde{\gamma}_{\ell}^{a} (\sigma_{k}^{[N]})^{-1} = \widetilde{\gamma}_{\ell}^{a} \qquad (\ell \neq k,\, k+1)\,,
\end{eqnarray}
without a minus sign.
Thus, the operator $\sigma_{k}^{[N]}$ acting on $\widetilde{\gamma}_{l}^{a}$ 
reproduces the transformation (\ref{eq:exchange_tilde}).
We note that $\sigma_{k}^{[N]}$ can be expressed in terms of
$\widetilde{\gamma}_{k}^{a}$ only. 
On the other hand, $\overline{\gamma}_{k}$ is transformed by the operator
$h_{k}^{[N]}$ like a single Majorana fermion as demonstrated in
Eqs.~(\ref{eq:h-1})-(\ref{eq:h-3}), and hence $h_{k}^{[N]}$ reproduces
the transformation (\ref{eq:exchange_bar}).

To summarize this subsection, in correspondence to the product of 
$\tau_{k}^{[N]}=\sigma_{k}^{[N]}h_{k}^{[N]}$, the Majorana
operator $\gamma_{k}^{a}$ is also expressed by the product of
the two parts, $\widetilde{\gamma}_{k}^{a}$ obeying the Coxeter group given
by $\sigma_{k}^{[N]}$ and $\overline{\gamma}_{k}$ obeying Ivanov's
exchange given by $h_{k}^{[N]}$.

\setcounter{equation}{0}
\section{Operator decomposition and matrix representation \label{sec:tensor}}

So far, we have been discussing the factorized structure of the 
exchange operation of two vortices at the operator level. Everything was 
written in terms of the Majorana operators, and the decomposition into the 
Coxeter and Ivanov parts was naturally understood by using the 
Majorana operators. In contrast, in order to obtain the matrix 
representation, the usual procedure is to 
define Dirac fermion operators and use them in constructing the Hilbert 
space. Since the Dirac fermion operators are constructed from two Majorana 
fermions located separately at different vortices, it is not trivial if the 
factorized structure at the operator level is preserved in the matrix 
representation. For example, the Dirac fermion operator defined in 
Eq.~(\ref{eq:Dirac}) can not be decomposed 
similarly as the Majorana fermion operator as shown in 
Eq.~(\ref{decomp_Majorana}). In this section, we are going 
to show that the decomposition holds even in the matrix representation
in a suitable basis, 
and discuss the relationship between the the decompositions in the operator- 
and matrix-representation levels.

In the following, we discuss the case with $N=3$ for simplicity, but 
present the procedures to obtain the matrix representation so that they 
can be easily extended to the case with any odd $N$.

\subsection{Construction of the Hilbert space}

Let us consider an even number $n=2m$ of vortices. Then we can 
construct $m$ Dirac fermion operators $\Psi^a_K \
(a=1,2,3;\ K=1,\cdots, m)$, given in Eq.~(\ref{eq:Dirac}), in the 
vector representation of $SO(3)$. The Fock vacuum $\ket{0}$ 
is defined by the Dirac fermion operators as
\begin{eqnarray}
 \Psi_{K}^a \ket{0} =0\, ,  \quad {\rm for\ all}\ K\ {\rm and}\ a\, .
\end{eqnarray} 
One can construct the basis of the Hilbert space by acting the Dirac 
fermion operators $\Psi_{K}^{a\dagger}$ on the Fock vacuum $\ket{0}$.
The explicit forms of the basis were given in Ref.~\cite{Yasui:2010yh} 
for $N=3$ and $n=2,4$. Now let us construct the Hilbert space in 
a way different from Ref.~\cite{Yasui:2010yh}.
To this end, we define the number operator 
for the triplet Dirac fermions of the $K$-th pair of vortices by
\begin{equation}
 {\N}_K^a \equiv \Psi_K^{a\dagger}\Psi_K^a. \label{eq:triplet-number}
\end{equation}
A generic state for the $K$-th pair of vortices 
can be expressed in terms of the eigenvalues of 
this number operator as
\begin{equation}
 \ket{\N^1_K, \N^2_K, \N^3_K}^{}_K \label{eq:k-th-state}
\end{equation} 
where $\N^a_K=0$ or $1$ is the occupation number of the fermion 
created by $\Psi^{a\+}_K$ (here we use the same character for the 
operator and its eigenvalues).
Then the basis of the whole Hilbert space is composed of 
the tensor product of the states for each $K$,
\begin{equation}
\bigg\{\bigotimes_{K=1}^m  \ket{\N^1_K, \N^2_K, \N^3_K}^{}_K\bigg\} .
\label{eq:Hilbert-basis}
\end{equation}

\subsection{Singlet Dirac operators}

When $N=3$, the singlet Majorana operators $\overline{\gamma}_{k}$ 
defined in Eq.~(\ref{eq:composite}) are given by 
\begin{equation}
 \overline{\gamma}_{k} = \frac{1}{3 !} i \epsilon^{abc} 
\gamma_k^{a}
\gamma_k^{b}
\gamma_k^{c}\, .
\end{equation}
By using these operators, we define singlet Dirac operators 
\begin{equation}
 \overline{\Psi}_K 
\equiv \frac{1}{2} 
\left(
\overline{\gamma}_{2K-1} + i 
\overline{\gamma}_{2K}
\right), \quad 
\overline{\Psi}^\+_K 
\equiv \frac{1}{2} 
\left(
\overline{\gamma}_{2K-1} - i 
\overline{\gamma}_{2K}
\right). \label{eq:singlet-Dirac}
\end{equation}
These operators play a particular role when they act on the states. 
Let us examine the action of the singlet Dirac operators 
$\overline{\Psi}_K$'s and $\overline{\Psi}^\+_K$'s 
on the Hilbert space (\ref{eq:Hilbert-basis}), 
which is constructed by acting the triplet Dirac operators 
$\Psi^{a \+ }_K$'s and $\Psi^a_K$'s on the Fock vacuum $\ket{0}$.
To this end, we express the singlet Dirac operators 
$\overline{\Psi}_K$ and $\overline{\Psi}^\+_K$ in terms of 
the triplet Dirac operators ${\Psi}^a_K$ and ${\Psi}^{a\+}_K$.
For that purpose, we note that the singlet 
Majorana operators $\overline{\gamma}_{2K-1}$ and $\overline{\gamma}_{2K}$ 
defined in Eq.~(\ref{eq:composite}) can be written in two ways:
\begin{equation}
\begin{split}
 \overline{\gamma}_{2K-1} 
&=\overline{\Psi}_K +  \overline{\Psi}^\+_K  \\
&= \frac{1}{3 !} i \epsilon^{abc} 
(\Psi_K^a + \Psi_K^{a\+} )
(\Psi_K^b + \Psi_K^{b\+} )
(\Psi_K^c + \Psi_K^{c\+} ), \\
\overline{\gamma}_{2K}
&= 
( \overline{\Psi}_K -  \overline{\Psi}^\+_K  )/i 
\\
&= 
 \frac{1}{3 !} i \epsilon^{abc} 
\left( \frac{1}{i} \right)^3
(\Psi^a_K - \Psi^{a\+}_K) 
(\Psi^b_K - \Psi^{b\+}_K) 
(\Psi^c_K - \Psi^{c\+}_K) ,
\\
\end{split}
\end{equation}
respectively. 
From these two relations, we can express 
the singlet Dirac operators 
$\overline{\Psi}_K$ and $\overline{\Psi}^\+_K$ 
in terms of the triplet Dirac operators $\Psi_K^a$ and $\Psi_K^{a\+}$ 
as 
\begin{eqnarray}
\overline{ \Psi}_K 
&=&
 \frac{1}{3 !} i \epsilon^{abc} 
\left(  
\Psi_K^a \Psi_K^b \Psi_K^{c\+}
+ 
\Psi_K^a \Psi_K^{b \+} \Psi_K^{c}
+
\Psi_K^{a \+} \Psi_K^{b } \Psi_K^{c}
+ 
\Psi_K^{a \+} \Psi_K^{b \+} \Psi_K^{c \+}
\right) ,  \nonumber \\
\overline{ \Psi}^\+_K 
&=& 
- \frac{1}{3 !} i \epsilon^{abc} 
\left(  
\Psi_K^{a\+} \Psi_K^{b\+} \Psi_K^{c}
+ 
\Psi_K^{a\+} \Psi_K^{b} \Psi_K^{c\+}
+
\Psi_K^{a} \Psi_K^{b\+ } \Psi_K^{c\+}
+ 
\Psi_K^{a} \Psi_K^{b} \Psi_K^{c}
\right) .
\label{eq:bar-psi}
\end{eqnarray}

For the later purpose, we define the fermion number operator 
for the $K$-th pair of vortices 
in terms of the singlet Dirac operators $\overline{\Psi}^\+_K$
and  $\overline{\Psi}_K$:
\begin{equation}
\overline{\N}_K 
\equiv 
\overline{ \Psi}^\+_K 
\overline{ \Psi}^{}_K  .
\label{eq:fermion-num}
\end{equation}
The meaning of this number operator will be clarified below.

Now we can discuss how the singlet Dirac operators
$\overline{\Psi}_K$ and $\overline{\Psi}^\+_K$ act on the
Fock states (\ref{eq:k-th-state}) 
defined by the triplet Dirac operators 
$\Psi^a_K$ and $\Psi^{a^\+}_K$. 
The action of the singlet Dirac operators
$\overline{\Psi}_K$ and $\overline{\Psi}^\+_K$ on
$\ket{\N^1_K,\N^2_K,\N^3_K}_K$ can be read off from 
Eqs.~(\ref{eq:bar-psi}) as
\begin{eqnarray}
&& \overline{\Psi}_K \ket{0,0,0}_K = \ket{1,1,1}_K, \nonumber\\
&& \overline{\Psi}_K \ket{1,0,0}_K = 
\overline{\Psi}_K \ket{0,1,0}_K =
\overline{\Psi}_K \ket{0,0,1}_K
= 0, \nonumber\\
&& \overline{\Psi}_K \ket{0,1,1}_K = \ket{1,0,0}_K, \quad 
   \overline{\Psi}_K \ket{1,0,1}_K = \ket{0,1,0}_K, \quad
   \overline{\Psi}_K \ket{1,1,0}_K = \ket{0,0,1}_K,   \nonumber\\
&& \overline{\Psi}_K \ket{1,1,1}_K = 0,  \label{eq:bar-psi-k}
\end{eqnarray}
for $\overline{\Psi}_K$, and  %
\begin{eqnarray}
&& \overline{\Psi}^\+_K \ket{0,0,0}_K = 0, \nonumber\\
&& \overline{\Psi}^\+_K \ket{0,0,1}_K = \ket{1,1,0}_K, \quad 
   \overline{\Psi}^\+_K \ket{0,1,0}_K = \ket{1,0,1}_K, \quad 
   \overline{\Psi}^\+_K \ket{1,0,0}_K = \ket{0,1,1}_K, \nonumber\\
&& \overline{\Psi}^\+_K \ket{0,1,1}_K = 
   \overline{\Psi}^\+_K \ket{1,0,1}_K = 
   \overline{\Psi}^\+_K \ket{1,1,0}_K = 0,  \nonumber\\
&& \overline{\Psi}^\+_K \ket{1,1,1}_K = \ket{0,0,0 }_K, \label{eq:bar-psi-k2}
\end{eqnarray}
for $\overline{\Psi}^\+_K$.

Let us define the total fermion number operator for 
the $K$-th pair of vortices as the sum of the number operator 
$\N^a_K$ in Eq.~(\ref{eq:triplet-number}) of the triplet Dirac fermions:
\begin{equation}
 {\cal F}_K \equiv \sum_{a=1}^N \N^a_K .
\end{equation}
Then we can deduce that the states annihilated by 
$\overline{\Psi}_K$ are those with odd eigenvalues 
of ${\cal F}_K$ (see the second and fourth lines of 
Eq.~(\ref{eq:bar-psi-k})), while those with even eigenvalues of 
${\cal F}_K$ are annihilated by $\overline{\Psi}^{\+}_K$ 
(see the first and third lines of Eq.~(\ref{eq:bar-psi-k2})):
\begin{equation}
\overline{\Psi}^\+_K \ket{\mathrm{even}}_K = 0, \quad 
\overline{\Psi}_K \ket{\mathrm{odd}}_K = 0, 
\label{eq:action-psibar}
\end{equation}
where $\ket{\mathrm{even}}_K$ and $\ket{\mathrm{odd}}_K$ are 
eigenstates of the parity operator
\begin{equation}
\mathcal{P}_K \equiv (-1)^{{\cal F}_K},
\end{equation} 
namely, 
\begin{equation}
 \mathcal{P}_K \ket{\mathrm{even}}_K = +1 \ket{\mathrm{even}}_K, 
\quad
 \mathcal{P}_K \ket{\mathrm{odd}}_K = -1 \ket{\mathrm{odd}}_K.
\end{equation}
It should also be noted that, when $\overline{\Psi}^\+_K$ and 
$\overline{\Psi}_K$
do not annihilate the state, they create a state with 
opposite parity: 
\begin{equation}
\overline{\Psi}^\+_K \ket{\mathrm{odd}}_K = \ket{\mathrm{even}}_K, \quad 
\overline{\Psi}_K \ket{\mathrm{even}}_K = \ket{\mathrm{odd}}_K.
\label{eq:action-psibar-2}
\end{equation}
These facts imply that the parity operator $ \mathcal{P}_K$ 
anti-commutes with  $\overline{\Psi}_K$ and $\overline{\Psi}^\+_K$:
\begin{equation}
\{ \mathcal{P}_K, \overline{\Psi}_K \} = 0, \quad 
\{ \mathcal{P}_K, \overline{\Psi}^\+_K \} = 0.
\end{equation}

We claim that the fermion number operator $\overline{\N}_K$ is 
related with the parity operator $\mathcal{P}_K$ as
\begin{equation}
 \mathcal{P}_K = 1 - \overline{\N}_K.
\label{eq:parity}
\end{equation}
Namely, the fermion number operator $\overline{\N}_K$ 
expresses the parity of the total fermion number for each index $K$.
This relation results from Eq.~(\ref{eq:action-psibar}), 
which is obvious from the structure of
the operators shown in Eqs.~(\ref{eq:bar-psi}).

By repeating the same argument, 
the above relation can be generalized to an arbitrary odd $N$ as 
\begin{eqnarray}
\mathcal{P}_K &=& 1 - \overline{\N}_K  \quad \quad\, \mathrm{for} 
\quad N = 4\ell+3, \\
\mathcal{P}_K &=& \overline{\N}_K \quad \quad \quad\quad\mathrm{for} 
\quad N = 4\ell+1 .
\end{eqnarray}

\subsection{The tensor product structure of the Hilbert space}

From Eqs.~(\ref{eq:bar-psi-k}) and (\ref{eq:bar-psi-k2}), 
we can say that the action of the singlet Dirac operators
$\overline{\Psi}_K$ or $\overline{\Psi}^\+_K$ is 
a kind of ``NOT'' operation, which ``flips'' the fermion number 
for each component of the $SO(3)$ vector. Namely, under the action of 
$\overline{\Psi}_K$, one finds $\ket{\N^1_K,\N_K^2,\N_K^3}~{}_K\, \rightarrow 
\ket{1-\N^1_K,1-\N_K^2,1-\N_K^3}^{}_K$.
We can divide the states into pairs, in each of which the two states are 
related by the NOT operation. When $N=3$, there are four pairs:
\begin{eqnarray}
&& \ket{0,0,0}_K \leftrightarrow \ket{1,1,1}_K, \qquad
   \ket{0,0,1}_K \leftrightarrow \ket{1,1,0}_K, \nonumber\\
&& \ket{0,1,1}_K \leftrightarrow \ket{1,0,0}_K, \qquad
   \ket{1,0,1}_K \leftrightarrow \ket{0,1,0}_K.
\label{eq:pairs}
\end{eqnarray}
It is important to note that the singlet Dirac operators 
$\overline{\Psi}_K$ and $\overline{\Psi}^\+_K$ induce the
transition only within these pairs when they act on states
(e.g. a transition between $\ket{0,0,0}_K $ and $ \ket{1,1,1}_K$), 
but they never
induce an inter-pair transition.
This fact is essential in the following discussion.

Now we are ready to discuss how the decomposition of the exchange 
operator $\tau_k^{[N]}$ 
results in the tensor-product structure in a matrix representation.
From the analysis above, we can take the basis of the Hilbert space 
which is labeled by the parity $\mathcal{P}_K$ of the number of fermions 
with index $K$, 
and an additional index $m_K$ which labels the choice of a pair 
in Eq.~(\ref{eq:pairs}). We denote the states by 
\begin{equation}
\ket{\mathcal{P}_K,m_K }_K \equiv \ket{\mathcal{P}_K}_K 
\otimes \ket{m_K}_K .
\end{equation}
Let us consider the matrix elements 
$
 _K \bra{\mathcal{P}_K,m_K} \tau_k^{[N]}
 \ket{\mathcal{P}^\prime_K,m^\prime_K}_K
$
of the exchange operator $\tau_k^{[N]}$ in this basis.
Now we show that these matrix elements can be 
written as the tensor product
of the Ivanov matrix and the Coxeter matrix: 
\begin{equation}
 _K \bra{\mathcal{P}_K,m_K} \tau_k^{[N]} 
\ket{\mathcal{P}_K^\prime, m_K^\prime}_K
 =
{}_K \bra{m_K} \sigma_k^{[N]} \ket{m_K^\prime}_K \cdot
 {}_K \bra{\mathcal{P}_K} h_k^{[N]} \ket{\mathcal{P}_K^\prime}_K.
\end{equation}
Namely, the Coxeter operator $\sigma^{[N]}_k$ acts only on 
$\ket{m_K}_K$, while the Ivanov operator $h^{[N]}_k$
acts only on $\ket{\mathcal{P}_K}_K$.
To prove this, we show the following two statements for any $k$ and $K$:
\begin{enumerate}
 \item[(i)] $h^{[N]}_k$ acts as an identity on
	    $\ket{m_K}_K$. \label{tensor-1}
 \item[(ii)] $\sigma^{[N]}_k$ acts as an identity on 
$\ket{\mathcal{P}_K}_K$. \label{tensor-2}
\end{enumerate}

First, we can see that the statement (i) is true in the following way.
The Ivanov operator  $h^{[N]}_k$ is written by 
the singlet Majorana operators $\overline{\gamma}_k$
(see Eq.~(\ref{eq:h_k^N})), and hence by the singlet Dirac 
operators $\overline{\Psi}_L$ and $\overline{\Psi}^\+_L$, 
see  Eq.~(\ref{eq:singlet-Dirac}). 
When the singlet Dirac operators $\overline{\Psi}_L$ or 
$\overline{\Psi}^\+_L$ act on states, the only state that 
can be created is the one in which the fermion numbers are flipped.
So, the action of the singlet Dirac operators 
$\overline{\Psi}_L$ or $\overline{\Psi}^\+_L$ never induce 
an inter-pair transition. Therefore, the Ivanov operator 
$h^{[N]}_k$ does not change the index $m_K$, 
which labels the pair in Eq.~(\ref{eq:pairs}).

Next, the statement (ii) is equivalent to the statement that 
the Coxeter operator $\sigma^{[N]}_k$ and the total fermion 
number operator $\overline{\N}_K$ commutes. 
The total fermion number operator $\overline{\N}_K$ 
is written by the singlet Majorana operators 
$\overline{\gamma}_{2K-1}$ and $\overline{\gamma}_{2K}$, 
which commute with the Coxeter operator $\sigma^{[N]}_k$ 
as in Eq.~(\ref{eq:commute}). We thus find that the Coxeter operator
$\sigma^{[N]}_k$ and the total fermion number operator 
 $\overline{\N}_K$ also commute.

From these discussions, we have shown that the factorization of 
the exchange operator $\tau^{[N]}_k$ of vortices 
into the Ivanov operator $h^{[N]}_k$ and 
the Coxeter operator $\sigma^{[N]}_k$ results in the tensor-product
structure in the matrix representation.

\setcounter{equation}{0}
\section{Summary and discussion \label{sec:summary}}

We have considered non-Abelian statistics of vortices, 
each of which has $N$ Majorana zero-energy states 
inside its core on which an $SO(N)$ symmetry acts.
We have investigated how the degenerate states induced by zero modes are
transformed under an exchange of neighboring vortices.
We have shown that, for an arbitrary odd $N$, 
the exchange operator $\tau_k^{[N]}$ defined in Eq.~(\ref{eq:exchange-op}), 
generating the exchange of two neighboring vortices, 
can be factorized into two parts $\tau_k^{[N]}=\sigma_k^{[N]}h_k^{[N]}$ 
as seen in Eq.~(\ref{eq:formula-n}).  
The part which is given by $h_k^{[N]}$ defined in Eq.~(\ref{eq:def-h})
is essentially equivalent to the exchange operator introduced by Ivanov.
If it is expressed 
in terms of the composite singlet Majorana operator 
$\overline \gamma_k$ defined in Eq.~(\ref{eq:composite}), 
then it has the same form as the exchange operator $\tau_k$ 
in the case of the single Majorana fermion.
The other operator $\sigma_k^{[N]}$ defined in 
Eq.~(\ref{eq:def-sigma}) is a new part.  
We have proven in Appendix \ref{sec:Proof} that 
they constitute the Coxeter group of the type $A_{2m-1}$ 
(the symmetric group $S_{2m}$) for $n=2m$ vortices.
We have also shown in Sec.~\ref{sec:tensor} that the
factorization of the exchange operators
results in the tensor-product structure in its matrix 
representation in a suitable basis.

The $SO(N)$ symmetry considered in this paper is the largest 
symmetry group in the presence of $N$ Majorana fermions. 
Whether the symmetry is $SO(N)$ or its subgroups depends on 
the details of the systems. 
For instance, a higher (pseudo-)spin $S$ representation of 
$SO(3)$ contains $2S+1$ Majorana fermions, 
but the symmetry acting on them does not have to be $SO(2S+1)$.
Also the representation does not have to be irreducible; 
for instance four Majorana fermions may be 
decomposed into one singlet and one triplet of $SO(3)$, 
but the symmetry group does not have to be $SO(4)$. 
It will be interesting to extend our results to general representations 
including reducible representations. 
An extension to general groups also remains as an interesting 
future problem to be explored.

For vortices with 
an even number $N$ of Majorana fermions with the $SO(N)$ symmetry, 
we have not found any meaningful factorization of the exchange operator
$\tau_k^{[N]}$ so far.
It remains as a future problem to identify the non-Abelian statistics 
for even $N$ Majorana fermions.
When the symmetry group inside the vortex core is restricted 
to the unitary subgroup $U(N/2) \subset SO(N)$, 
$N$ Majorana fermions can be rearranged into 
$N/2$ complex Dirac fermions in each vortex. 
In this case, the situation is rather different  
because Dirac fermions are locally defined, 
and we do not need to define Dirac fermions non-locally 
by using two spatially separated vortices.  
Nevertheless we found a rather different kind of non-Abelian statistics 
in the case of $N=2$ with the $U(1)$ symmetry \cite{Yasui:2011gk}
and $N=4$ with the $U(2)$ symmetry \cite{Yasui:2012zb}.
Identifying the statistics for the general even $N$ of 
$N/2$ Dirac fermions also remains as a future problem.

Finally, it will be important to look for actual condensed matter 
systems realizing multiple Majorana fermions in the vortex core, 
and to study an impact of our results 
on applications to topological quantum computations.

\acknowledgements

Y.~H. is supported by the Japan Society for the Promotion of Science for
Young Scientists. 
S.~Y. is supported by a Grant-in-Aid for
Scientific Research on Priority Areas ``Elucidation of New
Hadrons with a Variety of Flavors (E01: 21105006).''
M.~N. is supported in part by 
Grant-in Aid for Scientific Research (No. 23740198) 
and by the ``Topological Quantum Phenomena'' 
Grant-in Aid for Scientific Research 
on Innovative Areas (No. 23103515)  
from the Ministry of Education, Culture, Sports, Science and Technology 
(MEXT) of Japan.

\renewcommand{\theequation}{A\arabic{equation}}

\appendix

\setcounter{equation}{0}
\section{Proof of Coxeter relations for arbitrary odd $N$}
\label{sec:Proof}

In this appendix we give a proof that the operators $\sigma^{[N]}_k$ 
defined in Eq.~(\ref{eq:def-sigma}) satisfy
the Coxeter relations in Eqs.~(\ref{eq:Coxeter-A2}) and (\ref{eq:Coxeter-A3}).

\subsection{$(
\sigma^{[N]}_k 
\sigma^{[N]}_\ell
)^3 = 1 $ for $|k-\ell| = 1
$}

In this subsection we show Eq.~(\ref{eq:Coxeter-A2}).
Let us note that the cube of the product of $\tau_{k}^{[N]}$ and
$\tau_{k+1}^{[N]}$ is written as 
\begin{equation}
( \tau_{k}^{[N]} \tau_{k+1}^{[N]} )^3
= 
(
\sigma^{[N]}_k
h^{[N]}_{k}
\sigma^{[N]}_{k+1}
h^{[N]}_{k+1}
)^3.
\end{equation}
Since 
$\sigma^{[N]}_k $ commutes with  $h^{[N]}_k$
and  $h^{[N]}_{k+1}$, the relation above is written as
\begin{equation}
( \tau_{k}^{[N]} \tau_{k+1}^{[N]} )^3
=
(
 \sigma^{[N]}_k
\sigma^{[N]}_{k+1}
)^3
(
h^{[N]}_{k}
h^{[N]}_{k+1}
)^3.
\label{eq:cubed}
\end{equation}

The left hand side of Eq.~(\ref{eq:cubed}) is equal to $-1$, which can
be shown as follows.
\begin{equation}
\begin{split}
 ( \tau_{k}^{[N]} \tau_{k+1}^{[N]} )^3 &= 
\prod_{a=1}^{N}
\left\{
 \frac{1}{\sqrt{2}}
 (1+\Gamma^a_k) 
 \frac{1}{\sqrt{2}}
 (1+\Gamma^a_{k+1})
\right\}^3 \\
&= 
\prod_{a=1}^{N}
\left\{
\frac{1}{2}
\left(
1 + \Gamma^a_k + \Gamma^a_{k+1} + \Gamma^a_k \Gamma^a_{k+1}
\right)
\right\}^3 \\
&= 
\prod_{a=1}^{N}
\frac{1}{2}
\left(
-1 + \Gamma^a_k + \Gamma^a_{k+1} + \Gamma^a_k \Gamma^a_{k+1}
\right)
\frac{1}{2}
\left(
1 + \Gamma^a_k + \Gamma^a_{k+1} + \Gamma^a_k \Gamma^a_{k+1}
\right)
 \\
&=(-1)^{N} \\
&= -1,
\end{split}
\label{eq:cube-tau}
\end{equation}
where in the third line we have used the relation, 
\begin{equation}
\left\{
 \frac{1}{2}
\left(
1 + \Gamma^a_k + \Gamma^a_{k+1} + \Gamma^a_k \Gamma^a_{k+1}
\right)
\right\}^2
=  
\frac{1}{2}
\left(
-1 + \Gamma^a_k + \Gamma^a_{k+1} + \Gamma^a_k \Gamma^a_{k+1}
\right),
\end{equation}
which follows from the anticommuting property of $\Gamma^a_k $ and  $\Gamma^a_{k+1} $.
In the fourth line, we have again used the anticommuting
property.

On the other hand, we can also show that 
$(
h^{[N]}_{k}
h^{[N]}_{k+1}
)^3$ is equal to $-1$, as 
\begin{equation}
\begin{split}
 (
h^{[N]}_{k}
h^{[N]}_{k+1}
)^3
&= 
\left\{
 \frac{1}{2}
\left(
1+ \Gamma^{(N)}_k
+ \Gamma^{(N)}_{k+1}
+  \Gamma^{(N)}_k
 \Gamma^{(N)}_{k+1}
\right)
\right\}^3 \\
&=
 \frac{1}{2}
\left(
- 1+ \Gamma^{(N)}_k
+ \Gamma^{(N)}_{k+1}
+  \Gamma^{(N)}_k
 \Gamma^{(N)}_{k+1}
\right)
 \frac{1}{2}
\left(
 1+ \Gamma^{(N)}_k
+ \Gamma^{(N)}_{k+1}
+  \Gamma^{(N)}_k
 \Gamma^{(N)}_{k+1}
\right)
\\
&= -1,
\end{split}
\label{eq:cube-h}
\end{equation}
where we have used
\begin{equation}
\left\{  \frac{1}{2}
\left(
1+ \Gamma^{(N)}_k
+ \Gamma^{(N)}_{k+1}
+  \Gamma^{(N)}_k
 \Gamma^{(N)}_{k+1}
\right)
\right\}^2
  =  
 \frac{1}{2}
\left(
-1+ \Gamma^{(N)}_k
+ \Gamma^{(N)}_{k+1}
+  \Gamma^{(N)}_k
 \Gamma^{(N)}_{k+1}
\right).
\end{equation}
From Eqs.~(\ref{eq:cubed}), (\ref{eq:cube-tau}), and (\ref{eq:cube-h}), we
can conclude 
$ (\sigma^{[N]}_k \sigma^{[N]}_\ell )^3 = 1$ for $|k-\ell| = 1$.

\subsection{$(
\sigma^{[N]}_k 
\sigma^{[N]}_\ell 
)^2 = 1$ for $|k-\ell| > 1$ }
To prove Eq.~(\ref{eq:Coxeter-A3}), we first note that 
by squaring both sides of Eq.~(\ref{eq:formula-n}) and using the relation
\begin{equation}
\left( \frac{1}{\sqrt{2}} (1+\Gamma^a_k)
\right)^2 = \frac{1}{2}(1+ 2\Gamma^a_k - 1) = \Gamma^a_k,
\end{equation}
one finds
\begin{equation}
 \Gamma^{(N)}_k  = (\sigma^{[N]}_k )^2 
\;
 \Gamma^{(N)}_k.
\end{equation}
Then, by multiplying $(\Gamma^{(N)}_k)^{-1}$ on both sides
from right, we obtain 
\begin{equation}
 (\sigma^{[N]}_k )^2 = 1.
\end{equation}
It follows that 
$
 (
\sigma^{[N]}_k 
\sigma^{[N]}_\ell 
)^2 = 1 $ for $|k-\ell| > 1$
since $\sigma^{[N]}_k $ and $\sigma^{[N]}_\ell $  with $ |k-\ell|
> 1$ commute.

We thus have shown that the operators $\sigma^{[N]}_k$ satisfy 
the Coxeter relation of the type $A_{2m-1}$.

\newpage



\begin{thebibliography}{99}
\bibitem{Jackiw:1981ee} 
  R.~Jackiw and P.~Rossi,
Nucl.\ Phys.\ B {\bf 190}, 681 (1981).  

\bibitem{ReadGreen:00}
N.~Read and D.~Green,
Phys.\ Rev.\ B {\bf 61}, 10267 (2000)
[arXiv:cond-mat/9906453].

%
\bibitem{Ivanov:2001}
D.~A.~Ivanov, 
Phys.\ Rev.\ Lett.\ {\bf 86}, 268 (2001)
[arXiv:cond-mat/0005069 [cond-mat.supr-con]].

\bibitem{Volovik:99}
G.~E.~Volovik,
JETP Lett. {\bf 70}, 609--614 (1999)
[arXiv:cond-mat/9909426].


\bibitem{Kitaev:2006}
A.~Kitaev, 
Ann.\ Phys.\ {\bf 303}, 2 (2003)[arXiv:quant-ph/9707021]; 
ibid.\ {\bf 321}, 2 (2006)
[arXiv:cond-mat/0506438]
.

\bibitem{Nayak:2008zza}
For a review, see 
  C.~Nayak, S.~H.~Simon, A.~Stern, M.~Freedman and S.~Das Sarma,
  Rev.\ Mod.\ Phys.\  {\bf 80}, 1083 (2008)
[arXiv:0707.1889 [cond-mat.str-el]].

\bibitem{NAstatistics}
%
A.~Stern, F.~von~Oppen, and E.~Mariani,
Phys.\ Rev.\ B\ {\bf 70}, 205338 (2004)
[arXiv:cond-mat/0310273];
%
M.~Stone and S.~Chung, 
Phys. Rev. B 73, 014505 (2006)
[arXiv:cond-mat/0505515];
  M.~Sato,
Phys.\ Lett.\ B {\bf 575}, 126 (2003)
[arXiv:hep-th/0307005].  

\bibitem{SchnyderRFL:08}
A.~Schnyder, S.~Ryu, A.~Furusaki, and A.~Ludwig, 
Phys.\ Rev.\ B {\bf 78}, 195125 (2008); AIP Conf. Proc. {\bf 1134}, 10 (2009)
[arXiv:0803.2786 [cond-mat.mes-hall]]; 
%
A.~Kitaev, 
Proceedings of the L.D.Landau Memorial Conference
``Advances in Theoretical Physics,'' Chernogolovka, Moscow region,
Russia, 22-26 June 2008 (unpublished).
%
\bibitem{Roy:2010}
R.~Roy, 
Phys.\ Rev.\ Lett.\ {\bf 105}, 186401 (2010)
[arXiv:1001.2571 [cond-mat.supr-con]].

\bibitem{Teo:2009qv}
  J.~C.~Y.~Teo and C.~L.~Kane,
  Phys.\ Rev.\ Lett.\ {\bf 104}, 046401 (2010)
  [arXiv:0909.4741 [cond-mat.mes-hall]];
  J.~McGreevy and B.~Swingle,
Phys.\ Rev.\ D {\bf 84}, 065019 (2011)
 [arXiv:1106.0004 [hep-th]];  
  S.~-H.~Ho,
Phys.\ Rev.\ D {\bf 84}, 127701 (2011)
[arXiv:1106.2144 [hep-th]].  

\bibitem{Freedman:2010ak} 
  M.~Freedman, M.~B.~Hastings, C.~Nayak, X.~-L.~Qi, K.~Walker and Z.~Wang,
Phys.\ Rev.\ B {\bf 83}, 115132 (2011)
[arXiv:1005.0583 [cond-mat.mes-hall]].  

\bibitem{Yasui:2010yh}
  S.~Yasui, K.~Itakura, M.~Nitta,
  Phys.\ Rev.\  {\bf B83}, 134518 (2011)
  [arXiv:1010.3331 [cond-mat.mes-hall]].

\bibitem{Coxeter}
H.~S.~M.~Coxeter,
Ann. Of Math. {\bf 35}, 588-621 (1934); 
J. London Math. Soc. {\bf 10}, 21-25 (1935);
J.~E.~Humphreys, ``{\it Reflection Groups and Coxeter Groups},'' 
Cambridge studies in advanced mathematics, 29 (1990). 

\bibitem{Alford:2007xm}
  M.~G.~Alford, A.~Schmitt, K.~Rajagopal, T.~Schafer,
  Rev.\ Mod.\ Phys.\  {\bf 80}, 1455 (2008)
  [arXiv:0709.4635 [hep-ph]].

\bibitem{Balachandran:2005ev}
  A.~P.~Balachandran, S.~Digal and T.~Matsuura,
  Phys.\ Rev.\  D {\bf 73}, 074009 (2006)
  [arXiv:hep-ph/0509276].
  
\bibitem{Nakano:2007dr}
  E.~Nakano, M.~Nitta and T.~Matsuura,
  Phys.\ Rev.\  D {\bf 78}, 045002 (2008)
  [arXiv:0708.4096 [hep-ph]];
Prog.\ Theor.\ Phys.\ Suppl.\  {\bf 174}, 254 (2008)
 [arXiv:0805.4539 [hep-ph]].  

\bibitem{Eto:2009kg} 
  M.~Eto and M.~Nitta,
Phys.\ Rev.\ D {\bf 80}, 125007 (2009)
  [arXiv:0907.1278 [hep-ph]].  

\bibitem{Eto:2009bh}
  M.~Eto, E.~Nakano and M.~Nitta,
  Phys.\ Rev.\  D {\bf 80}, 125011 (2009)
  [arXiv:0908.4470 [hep-ph]].

\bibitem{Eto:2009tr}
  M.~Eto, M.~Nitta and N.~Yamamoto,
  Phys.\ Rev.\ Lett.\  {\bf 104}, 161601 (2010)
  [arXiv:0912.1352 [hep-ph]].

\bibitem{Hirono:2010gq}
  Y.~Hirono, T.~Kanazawa and M.~Nitta,
  Phys.\ Rev.\  D {\bf 83}, 085018 (2011)
  [arXiv:1012.6042 [hep-ph]].

\bibitem{Eto:2011mk}
  M.~Eto, M.~Nitta and N.~Yamamoto,
  Phys.\ Rev.\  D {\bf 83}, 085005 (2011)
  [arXiv:1101.2574 [hep-ph]];
  A.~Gorsky, M.~Shifman and A.~Yung,
Phys.\ Rev.\ D {\bf 83}, 085027 (2011)
 [arXiv:1101.1120 [hep-ph]].  

\bibitem{Yasui:2010yw}
  S.~Yasui, K.~Itakura and M.~Nitta,
  Phys.\ Rev.\  D {\bf 81}, 105003 (2010)
  [arXiv:1001.3730 [hep-ph]].

\bibitem{Fujiwara:2011za}
  T.~Fujiwara, T.~Fukui, M.~Nitta and S.~Yasui,
  Phys.\ Rev.\  D {\bf 84}, 076002 (2011)
  [arXiv:1105.2115 [hep-ph]].

\bibitem{Jackiw:2011tk} 
  R.~Jackiw and S.~-Y.~Pi,
  Phys.\ Rev.\  B {\bf 85}, 033102 (2012)
  [arXiv:1109.4580 [cond-mat.str-el]].

\bibitem{Yasui:2011gk}
  S.~Yasui, K.~Itakura and M.~Nitta,
  Nucl.\ Phys.\  B {\bf 859}, 261 (2012)
  [arXiv:1109.2755 [cond-mat.supr-con]].

\bibitem{Yasui:2012zb} 
  S.~Yasui, Y.~Hirono, K.~Itakura and M.~Nitta,
  arXiv:1204.1164 [cond-mat.supr-con].

\end{thebibliography}
\end{document}